# Automatic Prostate Zonal Segmentation Using Fully Convolutional Network with Feature Pyramid Attention

Yongkai Liu[1,2], Guang Yang[4], Sohrab Afshari Mirak[1], Melina Hosseiny[1], Afshin Azadikhah[1], Xinran Zhong[1], Robert E. Reiter[5], Yeejin Lee[1], Steven Raman[1,5], Kyunghyun Sung[1,2,3]

[1] Department of Radiological Sciences, David Geffen School of Medicine, University of California, Los Angeles, CA, USA
[2] Physics and Biology in Medicine IDP, David Geffen School of Medicine, University of California, Los Angeles, CA, USA
[3] Department of Bioengineering, David Geffen School of Medicine, University of California, Los Angeles, CA, USA
[4] National Heart and Lung Institute, Imperial College London, South Kensington, London, UK, SW7 2AZ
[5] Department of Urology, David Geffen School of Medicine, University of California, Los Angeles, CA, USA

Corresponding author: Yongkai Liu (e-mail: liuyongkai1009@g.ucla.edu).

This work is supported by funds from the Integrated Diagnostics Program, Department of Radiological Sciences & Pathology, David Geffen School of Medicine at UCLA.

**ABSTRACT** Our main objective is to develop a novel deep learning-based algorithm for automatic segmentation of prostate zone and to evaluate the proposed algorithm on an additional independent testing data in comparison with inter-reader consistency between two experts. With IRB approval and HIPAA compliance, we designed a novel convolutional neural network (CNN) for automatic segmentation of the prostatic transition zone (TZ) and peripheral zone (PZ) on T2-weighted (T2w) MRI. The total study cohort included 359 patients from two sources; 313 from a deidentified publicly available dataset (SPIE-AAPM-NCI PROSTATEX challenge) and 46 from a large U.S. tertiary referral center with 3T MRI (external testing dataset (ETD)). The TZ and PZ contours were manually annotated by research fellows, supervised by genitourinary (GU) radiologists. The model was developed using 250 patients and tested internally using the remaining 63 patients from the PROSTATEX (internal testing dataset (ITD)) and tested again (n=46) externally using the ETD. The Dice Similarity Coefficient (DSC) was used to evaluate the segmentation performance. DSCs for PZ and TZ were 0.74±0.08 and 0.86±0.07 in the ITD respectively. In the ETD, DSCs for PZ and TZ were 0.74±0.07 and 0.79±0.12, respectively. The inter-reader consistency (Expert 2 vs. Expert 1) were 0.71±0.13 (PZ) and 0.75±0.14 (TZ). This novel DL algorithm enabled automatic segmentation of PZ and TZ with high accuracy on both ITD and ETD without a performance difference for PZ and less than 10% TZ difference. In the ETD, the proposed method can be comparable to experts in the segmentation of prostate zones.

## I. INTRODUCTION

Prostate cancer (PCa) is the most common solid noncutaneous cancer in American men and the second most common causing cancer-related death [1]. Multiparametric MRI (mpMRI) has shown promising results for the detection and staging for clinically significant PCa [2], [3]. Previous studies have reported that PCa in transition and peripheral zones exhibit different morphological and functional characteristics on mpMRI. According to the guideline for prostate cancer diagnosis on mpMRI, the Prostate Imaging Reporting and Data System version 2.1 (PI-RADSv2.1) [4], [5], different MRI sequences are primarily used for giving a PI-RADS score to a PCa depending on whether the lesion arises in PZ or TZ [6]. A highly reproducible, automatic segmentation of prostate zones (ASPZ) can enable the consistent assignment of lesion location since manual segmentation of prostate zones is highly dependent on reader experience and expertise. Also, manual segmentation of prostate zones is usually time-consuming and ASPZ can help relieve clinician's cognitive workload [7].

Atlas based methods were previously proposed to segment the prostate zones [8]. Deep learning (DL) based methods, such as U-Net [9]and its variants [6], [10]–[13], have recently been developed to perform prostate AS. U-Net, an architecture based on fully convolutional networks (CNN), contains encoder and decoder sub-networks, where the encoder module is used to capture the higher semantic information, and the decoder module recovers spatial information. U-Net can classify pixels of the two zones and effectively localize and segment TZ and PZ. However, semantic information captured by U-Net may not be sufficient to describe the heterogeneous anatomic structures of the prostate and indiscernible borders between TZ and PZ, resulting in inconsistent and sub-optimal ASPZ performance.

In this study, we propose a new DL based method for automatic segmentation of prostate zones by developing a fully CNN with a novel feature pyramid attention mechanism. In particular, the proposed CNN consisted of three sub-networks, comprised of an improved deep residual network (based on the ResNet50) [14], a pyramid feature network with attention [15], and a decoder. We incorporated the ResNet50 to cope with heterogeneous prostate anatomy with high level semantic features and the pyramid network with attention is designed to capture information at multiple scales. The proposed DL model was evaluated using both internal testing dataset and external testing datasets on axial mpMRI slice. In addition, we compare the proposed method with inter-reader consistency using two independent expert based manual segmentations.

## II. MATERIALS AND METHODS

### A. MRI Datasets

With approval from the institutional review board (IRB), this study was carried out in compliance with the United States Health Insurance Portability and Accountability Act (HIPAA) of 1996. The MRI datasets were collected from two centers: 1) The Cancer Imaging Archive (TCIA) for SPIE-AAPM-NCI PROSTATEX (PROSTATEX) challenge [16] for development and internal testing of the model (n=250 and 63) and 2) data collected from a large, U.S. tertiary academic medical center with a highly curated mpMRI dataset with whole mount histopathology correlation for external testing of the model (n=46; age 45 to 73 years and weight 68 to 113 kg). Axial T2-weighted (T2) turbo spin-echo (TSE) slices (Table 1) were used for segmentation For the PROSTATEX data, both TZ and PZ were segmented in OsiriX (Pixmeo SARL, Bernex, Switzerland) by two MRI research fellows, where the contours were later cross-checked by both genitourinary (GU) radiologists (10-15 years of post-fellowship experience interpreting over 1,000 prostate mpMRI) and clinical research fellows. For the single institutional data, the pre-operative mpMRI scans performed on one of the three 3T MRI scanners (38 on Skyra, 1 on Prisma, and 7 on MAGNETON Vida; Siemens Healthcare, Erlangen, Germany) between October 2017 and December 2018 were included. Two clinical GU research fellows, supervised by GU radiologists, independently contoured TZ and PZ in a blinded fashion using OsiriX.

TABLE I
Detailed T2w TSE protocols used for two MRI datasets.

| Datasets | Internal Testing dataset (ITD) | External Testing dataset (ETD) |
|---|---|---|
| Spatial Resolution | 0.5x0.5x3.6mm$^3$ | 0.65x0.65x3.6mm$^3$ |
| Flip angle | 160° | 160° |
| Matrix Size | 380x380 | 320x320 |
| Field-of-View | 190x190 mm$^2$ | 208x208 mm$^2$ |
| Repetition Time/Echo Time | 5660 ms / 104 ms | 4000 ms / 109 ms |

### B. Proposed Deep Learning Model for Automatic Prostate Segmentation

The structure of the proposed fully convolutional network is shown in Figure 1. The network consists of three separate sub-networks, including the improved ResNet50 for the encoding of rich semantic information from original images, the feature pyramid attention network to help capture the information at multiple scales, and the naïve decoder network to recover the spatial information. The three sub-networks are connected to be an end-to-end prostate zonal segmentation pipeline.

The ResNet50 utilizes the skip connections to avoid vanishing gradients problems so that more convolutional layers can be added to the network. We improved the ResNet50 by removing the initial max pooling layer and using the regular block instead of the bottleneck block at stride 1 as the first block in the 4[th] layer, as shown in Fig. 1a. The dilated bottleneck block was employed as the second block in the 4[th] layer to remain the size of the receptive field. This can minimize any potential loss to the spatial information and alleviate the burden of the decoder.

The feature pyramid attention was added after the modified ResNet50 for better sensing the fine details at different scales (Fig. 1b). The 3x3, 5x5, 7x7 convolutions in the pyramid structure were used to extract features from different scales. The features from different scales were integrated progressively for more precise incorporation of adjacent scales of context features and then were multiplied by the features from the improved ResNet50 after a 1x1 convolution operation for the global prior attention. The output features will be then added with features from both the global pooling branch and the modified ResNet50.

The decoder network consisted of two convolutions and two upsampling layers to recover the image dimensions to the original size (Fig. 1c). The final output was fed into a multi-class soft-max classifier for simultaneous segmentation of TZ and PZ.

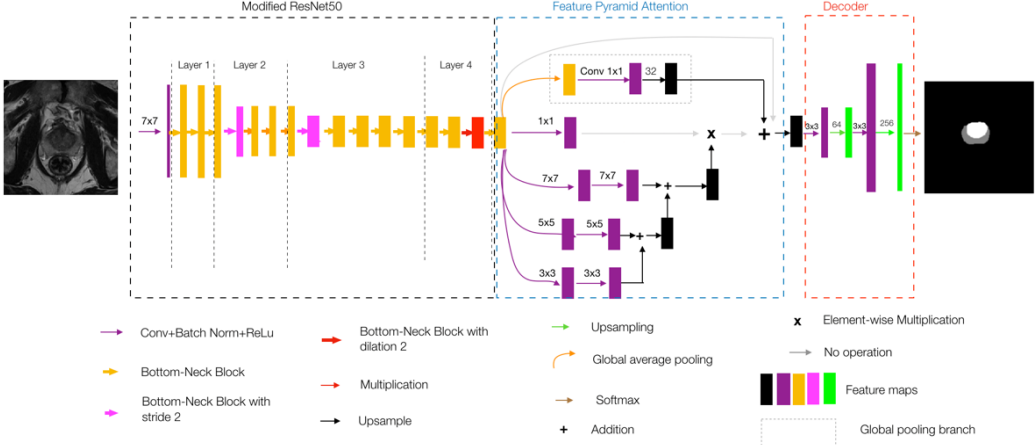

**FIGURE 1.** An overall structure of the proposed algorithm, where the input is a 2D slice of T2w MRI, and output is a mask showing the result of PZ and TZ segmentation (white – TZ and gray – PZ). The algorithm consists of three sub-networks - improved ResNet50 (a), feature pyramid attention (b), and decoder (c).

We used the cross entropy (CE) as the loss function for the proposed algorithm. For each given pixel, the cross entropy was defined as,

$$CE = \frac{1}{3}\sum_{i=0}^{3} -y_i \log(p_i) - (1 - y_i)\log(1 - p_i) \quad (1)$$

where $y_i \in \{0,1\}$ is the ground-truth binary indicator, corresponding to the 3-channel predicted probability vector $p_i \in [0,1]$. All the training and evaluation were performed on a desktop computer with a 64-Linux system with Titan Xp GPU with 12 GB GDDR5 RAM based on PyTorch. Learning rate was initially set to 2.5e-3, with momentum 0.9 and weight decay 0.0001. The model was trained for 100 epochs with batch size 48 and stochastic gradient descent. Since prostate areas are always in the middle, a central region (93mm x 93mm) was automatically cropped from original images before segmentation. Data augmentation methods were applied to increase the training data size, including flipped horizontally, rotated randomly between [-5°, 5°] and elastic transformations.

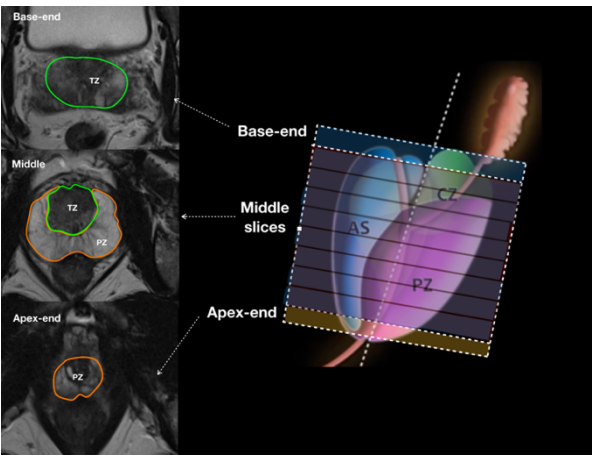

**FIGURE 2.** Representative examples of slices of prostate MRI. In left side, base-end slice (Only TZ exists), middle slice (both PZ and TZ exist) and apex-end slice (only PZ exists) are shown from top to bottom. The regions are encircled by green (TZ) and orange (PZ) boundaries.

### C. Model Development and Testing

A total of 250 patients' MRI from PROSTATEX were used for model development. Within the development dataset, 5-fold cross validation was adopted for model hyperparameter tuning and best model selection. The remaining 63 patients' MRI from PROSTATEX were used to test the model, which is called internal testing dataset (ITD) in this paper. A total of 47 patients' MRI from the large, U.S. tertiary academic medical center were served as another testing dataset, which is called external testing dataset (ETD).

For evaluation of the segmentation, we used the Dice Similarity Coefficient (DSC), formulated as:

$$DSC = \frac{2|X \cap Y|}{|X| + |Y|} \quad (2)$$

where X is the predicted 3D zonal segmentation on slices and Y is the ground-truth of 3D zonal contours on the slices.

From superior to inferior, we categorized prostate MRI slices into three levels, composed of base-end (includes mostly TZ), middle (includes mostly both TZ and PZ), and apex-end (includes mostly PZ), as shown in Figure 2. Both the prostate base-end and apex-end slices were identified when manual segmentation was performed, typically including one or two end slices of the prostate gland with only one prostate zone. A representative example of different prostate MRI slices is shown in Figure 3. DSCs were calculated considering different 3D zonal segmentation results, such as all slices (includes false positives), prostate slices (excludes false positives), base-end, middle, and apex-end slices. To assess the inter-reader consistency, we computed DSCs between two contours of TZ and PZ performed by two independent experts in a blinded fashion.

The corresponding imaging slices were used for the inter-reader agreement assessment.

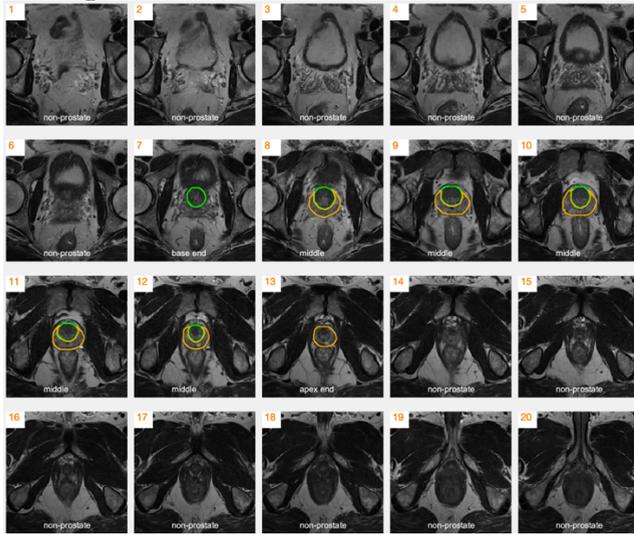

**FIGURE 3.** MRI slices from superior to inferior (slice 1 – 20). An example of non-prostate (slice 1-6, slice 14-20), base-end (slice 7), middle (slice 8-12) and apex-end (slice 13) slices is shown. Regions encircled by orange, green boundaries are PZ and TZ, respectively.

### D. Statistical analysis

Mean and standard deviation (SD) were used to summarize the distribution of DSCs. We performed the following three comparisons. First, we compared the performance of the proposed method with the baseline method – U-Net on the ITD dataset by using Wilcoxon rank-sum test. Second, we compared the performance of the proposed method on the ITD to the ETD by also using Wilcoxon rank-sum test. Third, we compared the performance of proposed method with the inter-reader agreement (Expert 1 vs. Expert 2) by using the Wilcoxon signed-rank test. P values less than 0.05 were considered statistically significant.

## III. RESULT

### A. Model Testing Using Internal Testing Dataset (ITD) and External Testing Dataset (ETD)

Two representative examples of automatic prostate zonal segmentation on ITD and ETD by our proposed method and U-Net are shown in Figure 4.

Table 2 includes mean and standard deviation of DSCs for PZ and TZ. Our proposed method achieved the mean DSC of 0.74 and 0.86 for PZ and TZ on ITD, mean DSC of 0.74 and 0.79 for PZ and TZ on ETD, which are all significantly larger than U-Net's results.

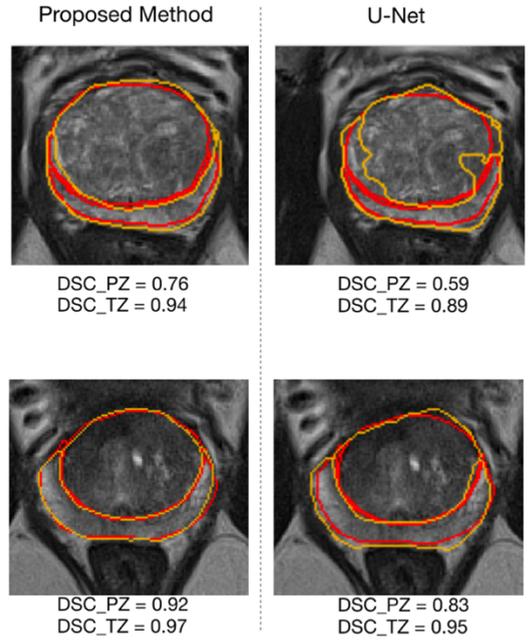

**FIGURE 4.** Representative examples of the automatic segmentation by the proposed method (orange lines) and U-Net in comparison with manual segmentation (red lines). DSCs are shown below the figures.

TABLE II
Performance of the proposed algorithm on internal testing dataset. *P* values are the comparisons between the proposed model's performance and the U-Net on internal testing dataset.

| Datasets | ITD | | ETD | |
|---|---|---|---|---|
| | PZ | TZ | PZ | TZ |
| U-Net | 0.69±0.10 | 0.83±0.09 | 0.67±0.09 | 0.76±0.10 |
| **Proposed Method** | 0.74±0.08 | 0.86±0.07 | 0.74±0.07 | 0.79±0.12 |
| | P<0.05 | P<0.05 | P<0.05 | P<0.05 |

Table III shows the performance of prostate zonal segmentation by the proposed model with Max-Pool and without Max-Pool on the ITD. After adding the Max-Pool in the ResNet50, mean DSCs for PZ and TZ are 0.72 and 0.84, which are smaller than the DSCs of proposed method (No Max-Pool in the ResNet50). This proves Max-Pool compromises the segmentation performance of prostate zones.

TABLE III
Performance comparison between the proposed model with Max-Pool and without Max-Pool under ITD. In our proposed method, we removed Max-Pool in the ResNet50.

| | Dice | |
|---|---|---|
| | PZ | TZ |
| Proposed Method | 0.74 ±0.08 | 0.86 ±0.07 |
| Add the Max-Pool in the proposed method | 0.72 ±0.08 | 0.84 ±0.07 |

## B. Comparison of Model performance on Internal Testing Dataset (ITD) and External Testing Dataset (ETD)

In Table IV, we show the performance of the proposed algorithm in the ITD and ETD. There was no significant difference of model's DSC between the ITD and the ETD for PZ.

However, for TZ, there was a small difference between model's DSC on the ITD and the ETD. The DSC differences of proposed algorithm on the ITD compared to the validation dataset were 8%.

TABLE IV
Performance of the proposed algorithm on ITD and the ETD. *P* values of model's performance on ITD relative to ETD are given and were obtained by using Wilcoxon rank-sum test.

| Datasets | PZ | TZ |
|---|---|---|
| ITD | 0.74 ±0.08  P=0.14 | 0.86 ±0.07  P<0.05 |
| ETD | 0.74 ±0.07 | 0.79 ±0.12 |

## C. Comparison Between Proposed Model and Experts under ETD

Examples of automatic segmentation results of slices by our proposed method, Expert 1 and Expert 2, at the base-end, middle, and apex-end on ETD are shown in Figure.5.

In Table 4 we show the DSCs of the proposed algorithm on different types of slices in the ETD and the inter-reader agreement between Expert 1 and Expert 2.

The proposed model's DSCs for both PZ and TZ are significantly larger than the inter-reader consistency in all slices, middle slices, apex-end slices and base-end slices when taking Expert 1's annotations as the ground-truth.

TABLE V
DSCs of the proposed algorithm on different types of slices in the external testing dataset. *P* values relative to inter-reader agreement (Expert 1 vs. Expert 2) are given in the Table for each and were obtained by using Wilcoxon Signed-Rank test.

| Comparisons | All slices | | Middle Slices | | Apex-End | | Base-End | |
|---|---|---|---|---|---|---|---|---|
| Zone | PZ | TZ | PZ | TZ | PZ | TZ | PZ | TZ |
| Model vs. Expert 1 | 0.74 ±0.07  P<0.05 | 0.79 ±0.12  P<0.05 | 0.75 ±0.07  P<0.05 | 0.83 ±0.09  P<0.05 | 0.84 ±0.11  P<0.05 | 0.77 ±0.21  P<0.05 | | |
| Expert 1 vs. | 0.71 ±0.13 | 0.75 ±0.14 | 0.71 ±0.13 | 0.81 ±0.12 | 0.76 ±0.21 | 0.65 ±0.27 | | |

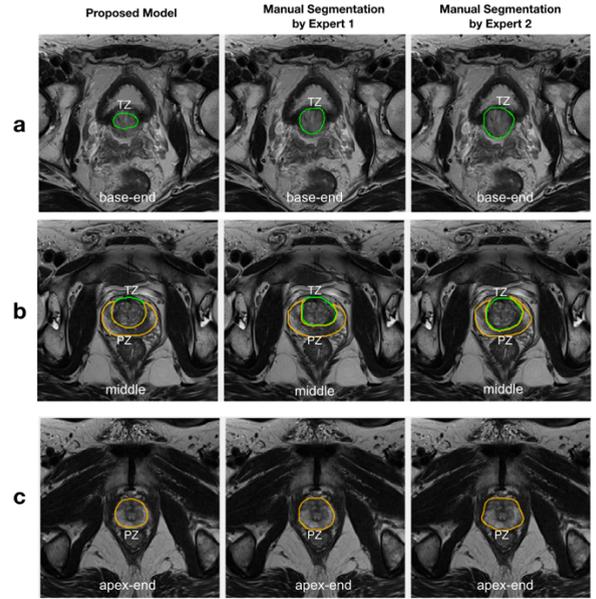

FIGURE 5. Representative examples of the automatic segmentation for testing, in comparison with manual segmentations by Expert 1 and 2. TZ is colored as green and PZ is colored as orange. From superior to inferior, base-end (a), middle (b), and apex-end (c) slices are shown with segmentations of the prostate zones.

## IV. DISCUSSION

We proposed a novel fully convolutional network-based model with feature pyramid attention for the automatic segmentation of the two prostate zones. The study showed that the proposed model performed consistently on ITD and ETD.

We observed slight differences between ITD and ETD, particularly in segmenting TZ. We believe this can be potentially due to 1) the differences of the imaging sequences, such as in-plane resolution and T2 contrast, 2) the discrepancies of the zonal annotations as different experts independently segmented the prostate zones for ITD and ETD.

We also found that the manual PZ segmentation was less consistent than the manual TZ segmentation, measured by DSCs between two experts (Table V). This may be due to the more complex structure of PZ as its boundaries are sometimes poorly discerned due to a variety of factors such as prostate or patient motion. Similarly, Meyer et. al, [7] reported that the PZ segmentation had worse inter-reader consistency than TZ segmentation with three different experts (first urologist, second urologist with the help of a medical student, and an assistant radiologist). Meyer et al. also utilized three orthogonal planes of the T2w MRI, i.e., sagittal, coronal and axial, to automatically determine the bounding box for the prostate before performing the segmentation. The bounding box approach could be added as pre-processing to improve both the segmentation performance and inter-reader consistency by minimizing the false positives.

In the ETD, when only considering middle slices for the testing, mean DSCs are higher than considering all slices. This may be because: 1) the features for the differentiation of PZ and TZ are more distinct in the middle slices than the other slices. 2) when we only consider the middle slices, some false positives from the non-prostate slices, apex-end slices and base-end slices can be avoided. Besides, we also found mean PZ DSC for apex-end slices is larger than the PZ DSC for middle slices, but to the contrary, TZ DSC for base-end slices is smaller than the TZ DSC for middle slices. The large standard deviations and low DSC of TZ for base-end slices indicated the significant discrepancies between two experts at the base-end slices. This indicates that it's hard to recognize TZ in the base-end slices, which may explain why the proposed method got a low TZ DSC in the base-end slices.

Compared with the DSCs of method of Meyer et al. [7], our method's DSCs for both PZ and TZ are lower. The reasons may be as follows: 1) Difference in sample sizes for the evaluation. In our method, 63 patient datasets were used for the testing data set, in compared with their testing data set of only 20 patients. 2) Discrepancy in manual annotations for both PZ and TZ. 3) Inherent differences in methods. 4) Differences of preprocessing. In their method, before the segmentation, the bounding box for the prostate was determined to reduce the false positives.

Our study also has a few limitations. Firstly, the vendor for the ITD is same with the ETD. Also, in-plane resolution of the ITD is very close to that of the ETD. Datasets from different vendors and with considerable different in-plane resolutions will be incorporated into future related studies. Secondly, the proposed algorithm is a 2D-based FCN model, which is still deficient in capturing inter-slice correlation information compared to 3D-based models. In the future, we will explore ways of improving the capturing of inter-slice correlation information in our proposed model. Thirdly, the number of experts involved in obtaining inter-reader consistency in the paper is two. In the future, more experts will be added in the study to get more robust inter-reader consistency.

## V. CONCLUSION

In this study, we proposed a novel deep learning algorithm for the automatic segmentation of the two prostate zones using T2w MRI. The proposed algorithm outperforms the U-Net on automatic segmentation of PZ and TZ. Besides, the difference between the proposed method's performance on internal testing dataset and external testing dataset is subtle, especially for the segmentation of PZ. Moreover, the performance of the proposed method is comparable to the experts in the external testing dataset.